\documentclass[preprint,letter,preprintnumbers,amsmath,amssymb,floatfix]{revtex4}

\usepackage{graphicx}
\usepackage{dcolumn}
\usepackage{bm}
\usepackage{slashed}
\usepackage{amssymb}

\newcommand{\beqa}{\begin{eqnarray}}
\newcommand{\eeqa}{\end{eqnarray}}
\newcommand{\be}{\begin{equation}}
\newcommand{\ee}{\end{equation}}
\newcommand{\nn}{\nonumber\\}
\def\Tr{\mbox{Tr}}
\def\Im{\mbox{Im}}

\begin{document}

\title{Electromagnetic Radiation in Hot QCD Matter:\\
Rates, Electric Conductivity, Flavor Susceptibility and Diffusion.}

\author{Chang-Hwan Lee$^{a,b}$ and Ismail Zahed$^b$}

\affiliation{(a) Department of Physics, Pusan National University, Busan 609-735, South Korea,\\
(b)~Department~of~Physics~and~Astronomy~Stony Brook University, NY 11794, USA}

\date{\today}

\begin{abstract}
We discuss the general features of the electromagnetic radiation from a thermal
hadronic gas as constrained by chiral symmetry. The medium effects on the 
electromagnetic spectral functions and the partial restoration of chiral 
symmetry are quantified in terms of the pion   densities. 
The results are compared with the electromagnetic radiation from a strongly interacting quark-gluon plasma in terms of the leading gluon condensate operators. We use the spectral functions as constrained by the emission rates
to estimate the electric conductivity, the  light flavor susceptibility and diffusion constant 
across the transition from the correlated hadronic gas to a strongly interacting quark-gluon plasma.
\end{abstract}

\maketitle

\section{Introduction}
One of the chief objectives of the ultra-relativistc heavy ion program at RHIC and LHC
is to excite enough of the QCD vacuum in the form of a quark-gluon plasma. The plasma
expands and hadronize relatively quickly making its identification only implicit through
the high hadronic multiplicities or electromagnetic emissivities \cite{Ada12,Geu13,Loh12,Hee11,Rap12,Rap13,Rap13b,She13,She13b,Vuj13}.  

Electromagnetic emissions in the form of dileptons or photons occur throughout the
life-time of the expanding fire-ball. The early stages are dominated by the emission from
the partonic constituents, while the late stages of the emission are dominated by the
hadronic constituents. Both the early and late stages are well-described by a hydro-dynamical
fire-ball. 
In this letter, instead of integrating over the space-time of the evolving fire-ball, we discuss the basics of the electromagnetic emissivities from a hadronic gas \cite{Ste96,Ste97,Lee98,Dus07,Dus09,Dus10} and a strongly
coupled plasma (sQGP) which is described in terms of Born diagrams \cite{McL85} corrected by
leading order gluon condensates \cite{Han87,Lee99,Bas14}. The comparison with the newly reported lattice
simulations of the electromagnetic spectral functions at zero momentum puts some
constraints on the importance on the gluon condensates \cite{Din11,Kac13}.

Dilepton and photon emissions are the results of  many 
reaction processes involving the quark-gluon plasma in the early stage and hadrons and the strong character of their interactions in the later stage.  
For the emissions from the hadronic gas, the only organizational principles are broken chiral symmetry and gauge invariance, both of which are difficult to assert  in individual reaction processes. In the spectral analysis \cite{Dey90,Hua95},
if hadrons thermalize with the pions and nucleons as the only strongly stable constituents,  there
is a way to systematically organize the electromagnetic emissivities by expanding them
not in terms of processes but rather in terms of final hadronic states.  
Then the emissivities from the hadronic gas can be represented by spectral functions by chiral reduction\cite{Ste96,Ste97}. These spectral functions
are either tractable from other experiments or amenable to resonance saturation\cite{Dus09b}. The spectral analysis allows us to represent the partial chiral symmetry restoration in terms of the mixing between vector and axial correlators.

In section~\ref{sec:had}, we review the spectral function approach to the photon and dilepton
rates emphasizing  the nature of the dynamical restoration of
the partially broken chiral symmetry in the hadronic fire-ball through the mixing of vector and axial correlators.
We also discuss the electric conductivity and the quark number susceptibilities in the correlated hadronic gas
near the chiral transition. In section~\ref{sec:sQGP} we review the sQGP corrected by the soft electric and magnetic
condensates and show that they may enhance the soft photon and dilepton emissions.  The electric conductivity
and the flavour diffusion constant in the sQGP are derived and compared to current lattice data.
Our conclusions are in section~\ref{sec:con}.

\section{Electromagnetic Radiation from Hadronic Gas}
\label{sec:had}

\subsection{Dilepton and Photon Rates }

In this section we review the spectral approach for the dilepton and photon production from a hadronic gas in thermal equilibrium \cite{Hua95,Ste96,Dus10}. The main advantage of the spectral function approach is that the calculation can be organized in a virial-like expansion and in principle all possible reaction channels can be included in the zero temperature spectral densities. The dilepton rate $R$, the number of dileptons produced per unit four  volume,  can be expressed using the current-current correlator as
\begin{eqnarray}
\label{ee}
\frac{dR}{d^4q}=\frac{-\alpha^2}{6\pi^3 q^2}
\,\left(1+\frac{2m^2_l}{q^2}\right)\left(1-\frac{4m^2_l}{q^2}\right)^{1/2}
{\bf W}(q)
\label{eq:dRdq}
\end{eqnarray}
where $\alpha=e^2/4\pi$ is the fine structure constant, $M \equiv \sqrt{q^2}$ is the dilepton invariant mass, $m_l$ is the lepton mass and the un-ordered 
electromagnetic current-current correlator  is given by \cite{Wel90,Ste97}
\beqa
{\bf W}(q)=
\int d^4x e^{-iq\cdot x}\Tr\left[ e^{-({\bf H}-{\bf F})/T} \,{\bf J}^\mu(x) {\bf J}_\mu(0)\right]\,.
\label{eq:cor}
\eeqa
Here ${\bf H}$ is the hadronic Hamiltonian,
${\bf F}$ is the Helmholtz free energy, $T$ is the temperature and $e {\bf J}_\mu$ is the hadronic part of the electromagnetic current,
\begin{eqnarray}
\label{eeJ}
{\bf J}_\mu(x)=\sum_{f} {\tilde e}_f\,\overline{\bf q}_f\gamma_\mu {\bf q}_f (x)
\end{eqnarray}
with ${\tilde e}_f=(2/3,-1/3,-1/3)$. Note that we consider only three flavors which will be valid for the thermal electromagnetic emission below the charmonium peak.

Using the un-ordered correlator, Eq.~(\ref{eq:cor}), the number of real photons produced per unit volume and unit three momentum 
can also be obtained as
\beqa
\frac{q^0 dN}{d^3q}=-\frac{\alpha}{4\pi^2}\,{\bf W}(q) 
\label{eq:photon}
\eeqa
with $q^2=0$. This equation with Eq.~(\ref{eq:dRdq}) enables us to link the {\it quasireal} virtual photon rate $N^*$ with dielectron data in the low mass  region below two pion threshold \cite{Ada10a,Ada10,Ada12}, 
\begin{equation}
\label{eeg}
 \frac{dR}{d^4q}  =  \frac{2\alpha}{3\pi M^2}
\,\left(1+\frac{2m^2_l}{M^2}\right)\left(1-\frac{4m^2_l}{M^2}\right)^{1/2}
\,\left(\frac {q^0 dN^*}{d^3q}\right)  .
\end{equation}
In the limit of ${M\rightarrow 0}$, $N^* \approx N$.

Symmetry and spectral analysis allows us to re-express the un-ordered
correlator in terms of the absorptive part of the Feynman  correlator\cite{Abr75},
\beqa
{\bf W}(q)=\frac{2}{e^{q^0/T}+1}\,{\rm Im}{\bf W}^F(q) 
\label{XX1}
\eeqa
where the Feynman correlator with time-ordering ($T^*$) is given by
\begin{equation}
{\bf W}^F(q)=i\int d^4x e^{iq\cdot x}\Tr\left[ e^{-({\bf H}-{\bf F})/T} {T}^*{\bf J}^\mu(x) {\bf J}_\mu(0)\right] \,.
\end{equation}
One can also obtain the retarded correlator from the Feynman correlator \cite{Abr75}
\begin{eqnarray}
\text{Im}{\bf W}^R(q)=\text{tanh}\left(q^0/2T\right)\text{Im}{\bf W}^F(q) .
\label{eer}
\end{eqnarray}
Using the retarded correlator one can obtain the electric conductivity from the linear response theory as we discuss later\cite{Din11}.

\subsection{Mixing of Vector and Axial Correlators in Pionic Gas} 
 
In Steele et al. \cite{Ste97} pion and nucleon contributions to the Feynman correlator were obtained within the context of a density expansion. For the heavy ion collisions where the net nucleon density is not negligible both pion and nucleon contributions are important \cite{Dus10}. However, for high energy collisions at RHIC and LHC, the pion contribution will dominate because the net baryon density of the fire ball becomes negligible. In this work, we focus on the pion contributions.
By taking the pion density as an expansion parameter, the pion contributions to the Feynman correlator can be expressed as
\begin{equation}
{\bf W}^F(q)={\bf W}^F_0 (q) + \frac{1}{f_\pi^2} \int d\pi {\bf W}^F_\pi(q,k)
+ \frac{1}{2!} \frac{1}{f_\pi^4} \int d\pi_1 d\pi_2 {\bf W}^F_{\pi\pi} (q,k_1,k_2)
+  \cdots
\label{eq:Fcor}
\end{equation}
where  
\beqa
{\bf W}^F_0(q) &=&i\int d^4x e^{iq\cdot x}\langle 0\vert T^* {\bf J}^\mu(x) {\bf J}_\mu(0)\vert 0\rangle \nn
{\bf W}^F_\pi(q,k) &=&i f_\pi^2 \int d^4x e^{iq\cdot x}\langle \pi^a(k)\vert T^* {\bf J}^\mu(x) {\bf J}_\mu(0)\vert \pi^a(k)\rangle \nn
{\bf W}^F_{\pi\pi}(q,k_1,k_2) &=&i f_\pi^4 \int d^4x e^{iq\cdot x}\langle \pi^a(k_1)\pi^b(k_2)\vert T^* {\bf J}^\mu(x) {\bf J}_\mu(0)\vert \pi^a(k_1)\pi^b(k_2)\rangle 
\label{WWW}
\eeqa
and    
\beqa
\int d\pi = \int \frac{d^3k}{(2\pi)^3 } \frac{n(E-\mu_\pi)}{2E}  
\eeqa
with $E = \sqrt{k^2 + m_\pi^2}$ and $n(\omega)=1/(e^{\omega/T}-1)$. Note that the finite pion chemical potential $\mu_\pi$ and the isospin sum over index $a$ and $b$ are included.

The first contribution ${\bf W}^F_0$ in (\ref{WWW}) is dominated by ${\bf \Pi}_V$, the transverse part of the vector correlator $\langle 0 | T^* {\bf VV } | 0\rangle$, which can be fixed by the measured electroproduction data\cite{Hua95,Lee98},
\be
{\rm Im} {\bf W}^F_0  = -3 \, q^2\,  {\rm Im} {\bf \Pi}_V (q^2) .
\label{eq:F0}
\ee  
This term vanishes for real photons with $q^2=0$ because the hadronic gas in thermal equilibrium is
stable against spontaneous photon emission.  
One pion contribution ${\bf W}^F_\pi$ can be represented by the measurable vacuum correlators using the
chiral reduction formulae~\cite{Ste96,Ste97},
\beqa
{\rm Im}{\bf W}^F_\pi(q,k)&=& 12 \, q^2\, \text{Im} {\bf \Pi}_V(q^2)\nn
& - &  6 \, (k+q)^2\text{Im} {\bf \Pi}_A \left( (k+q)^2\right) + (q\to -q)\nn
& + &  8 \left( (k\cdot q)^2-m_\pi^2 q^2\right) \text{Im} {\bf \Pi}_V(q^2)\times\text{Re} \Delta_R(k+q)+(q\to-q)
\label{eq:onepion}
\eeqa
where $\text{Re}\Delta_R = {\rm PP}\left[ 1/(k^2-m_\pi^2 + i\epsilon^2) \right]$ is the real part (principle value) of the retarded pion propagator
and ${\bf \Pi}_A$ is the transverse parts of the axial correlator $\langle 0 | AA | 0 \rangle$ which also can be fixed using experimental data \cite{Hua95,Lee98}. The full expression for the two pion contribution  is more complicated \cite{Ste97,Dus10} and the important contributions to ${\Im} {\bf W}_{\pi\pi}^F$ are summarized in  Appendix A.

\begin{figure}
\includegraphics[scale=.45]{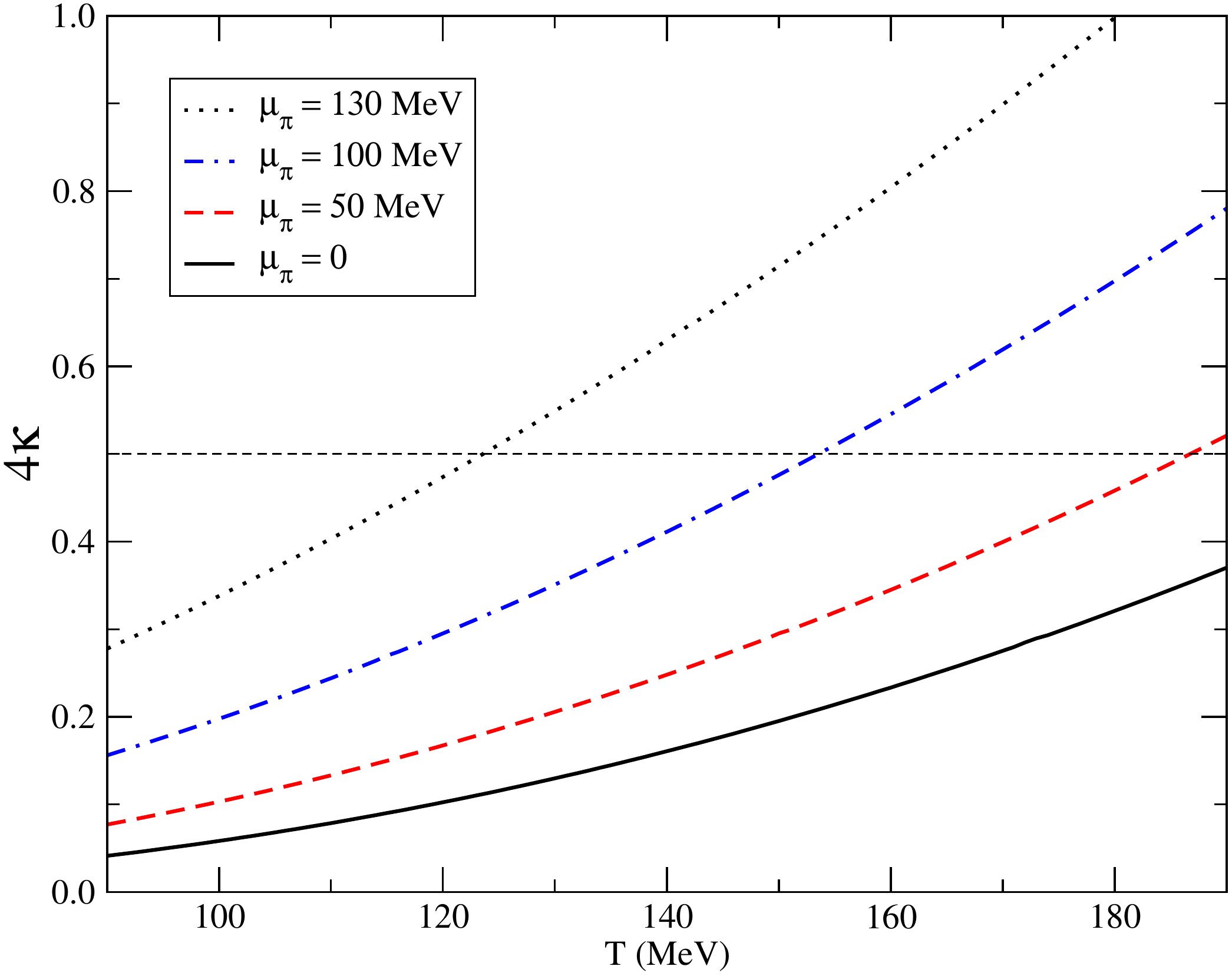}
\caption{Pion Density parameter $\kappa$ vs temperature for different $\mu_\pi$}
\label{kappa}
\end{figure}

\begin{figure}
\includegraphics[scale=.6]{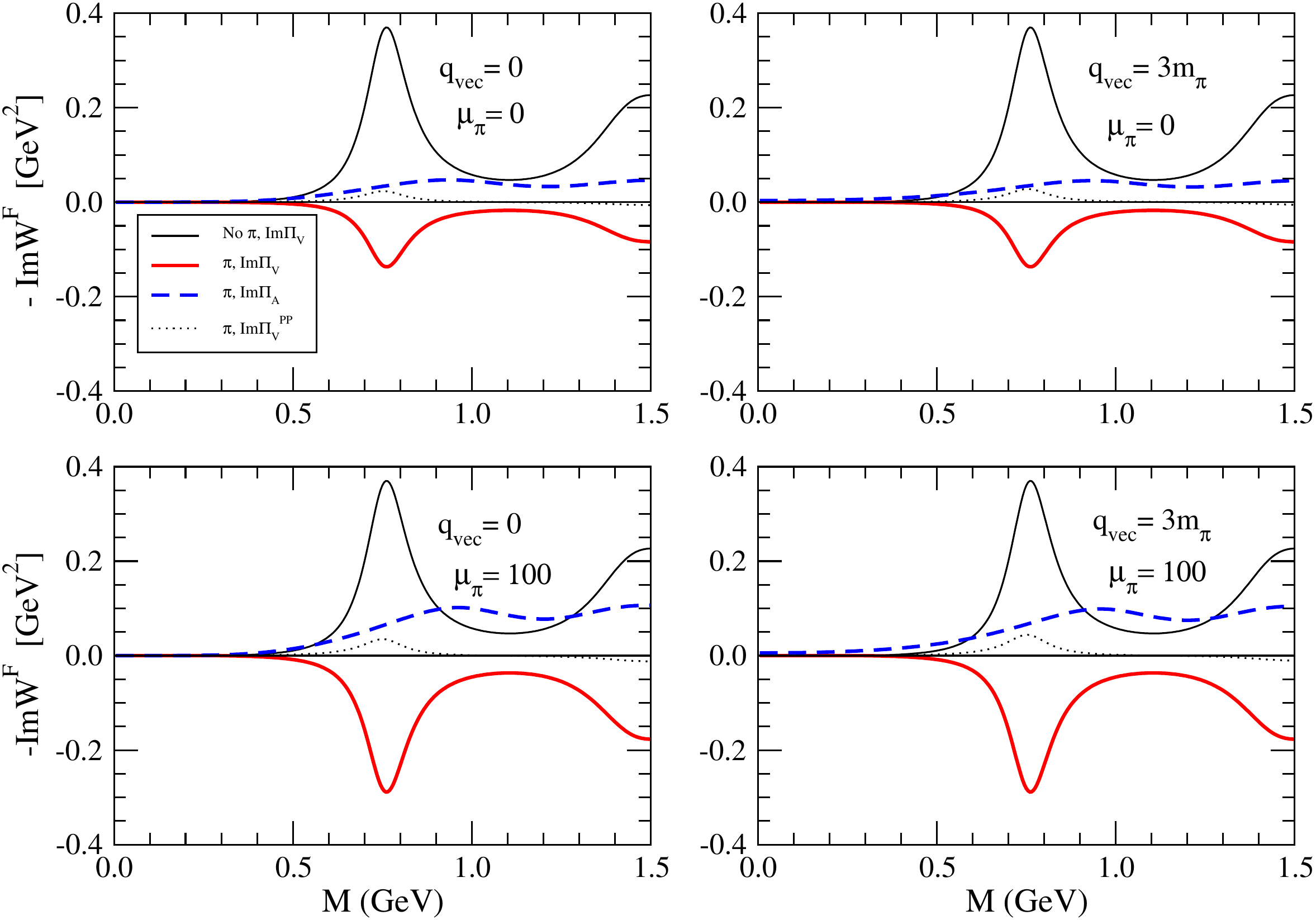}
\caption{Partial contributions of Eqs.~(\ref{eq:F0})  and (\ref{eq:onepion}) to the imaginary part of the correlator $-{\rm Im}{\bf W}^F$  at $T=190$ MeV  for different $|\vec q |$ and $\mu_\pi$. The thick black solid lines are the 0th order contribution without the pion. For the one pion contribution, labeled by $\pi$, the three lines in each figure correspond to the three lines  in Eq.~(\ref{eq:onepion}), respectively. PP represents the contribution from terms with the retarded pion propagator.}
\label{kappa2}
\end{figure}

\begin{figure}
\includegraphics[scale=.6]{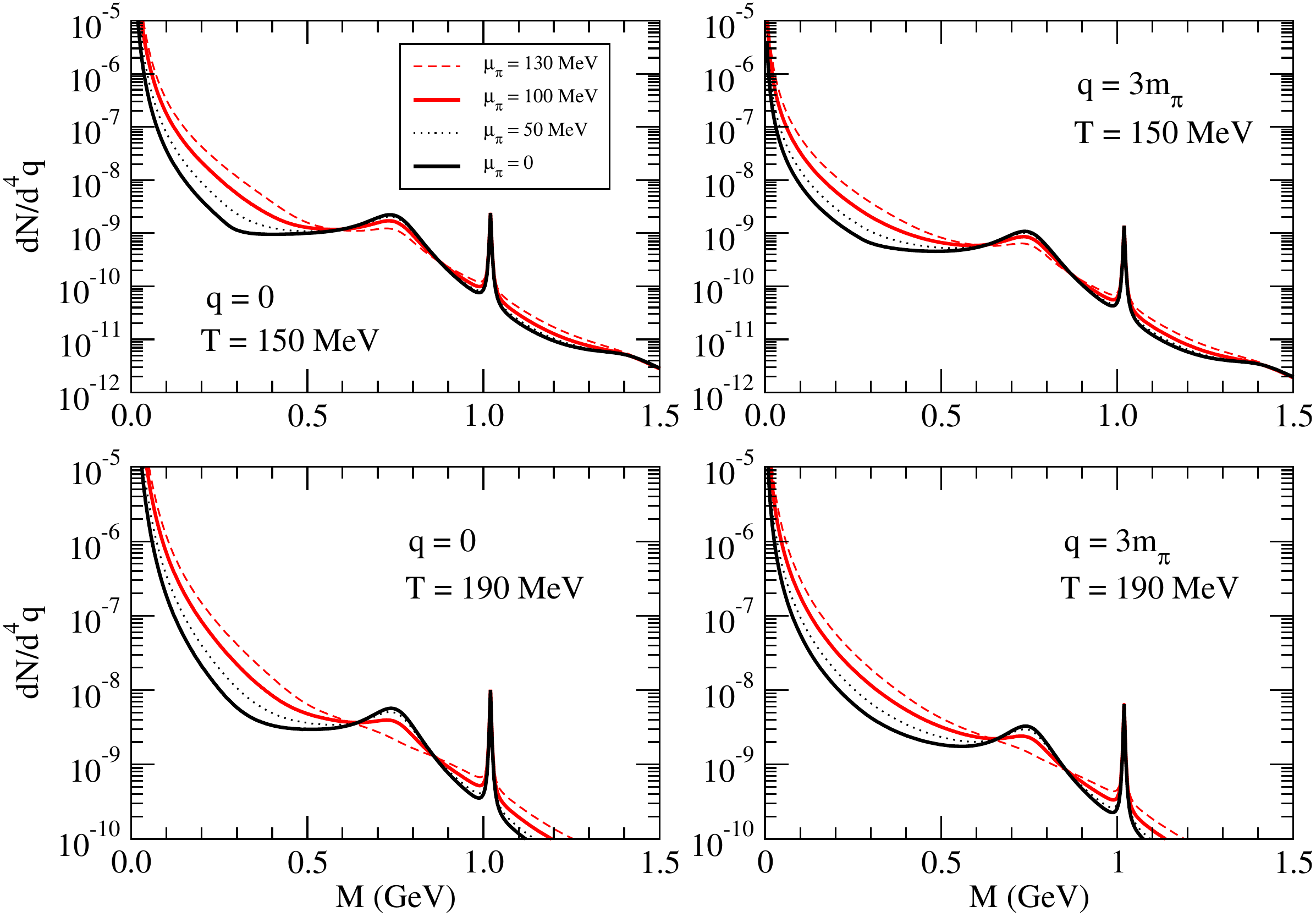}
\caption{Dilepton rates: Hadrons at $T= 150$ \& 190 MeV for various $|\vec{q}|$ and $\mu_\pi$}
\label{fig_dNdq_had}
\end{figure}

The mixing of vector and axial correlators as an indication of chiral symmetry restoration has been discussed in the literature in the limit of zero chemical potential and zero pion mass \cite{Dey90,Hua95}. In this work we extend the discussion in the presence of finite pion chemical potential and pion mass.  The pion density plays a major role for the mixing between the vector-axial correlators. 
In order to see the main idea of mixing, we focus on the contributions up to leading order in pion density. 
Firstly,  if we take $k\rightarrow 0$ and $m_\pi \rightarrow 0$ before the integration over the pion momentum \cite{Dey90,Hua95}, one can have a very schematic relation 
\beqa
{\rm Im} {\bf W}^F(q) & \approx & -3\, q^2 \left [ (1-4\kappa)\,{\rm Im}{\bf \Pi}_V(q^2)+4\kappa\,{\rm Im}{\bf \Pi}_A(q^2) \right] 
\label{eq:mix}
\eeqa
where $\kappa$ is the dimensionless pion phase-space factor  
\beqa
\kappa  = \frac{1}{f_\pi^2} \int d\pi \,\,.
\eeqa
The mixing is maximum for $\kappa\approx 1/8$, leading to the equal contribution from vector and axial correlators
\beqa
{\rm Im} {\bf W}^F(q) \propto
  {\rm Im}\left( {\bf \Pi}_V(q^2)+{\bf \Pi}_A(q^2)\right)  .
\label{zero1}
\eeqa

In Fig.~\ref{kappa} we show the dependence of $\kappa$ on the
temperature for different pion chemical potentials $\mu_\pi$. 
The vector-axial mixing (\ref{eq:mix}) is
enhanced at high temperature and/or higher $\mu_\pi$ as $\kappa$ increases. 
With the full expression, since Eq.~(\ref{eq:onepion}) depends on the pion momentum, the dependence on $\kappa$ is not-trivial.
In Fig.~\ref{kappa2} we show the partial contributions of Eqs.~(\ref{eq:F0})  and (\ref{eq:onepion}) to the imaginary part of the correlator, $- {\rm Im}{\bf W}^F$. In this figure one can clearly see that the one-pion contributions becomes significant as the pion chemical potential increases. The 50-50 mixing schematized in (\ref{zero1}) is apparent qualitatively at $\mu_\pi=100$ MeV with which there is a large cancellation among the contributions with ${\rm Im} {\bf \Pi}_V$.
In Fig.~\ref{fig_dNdq_had} the dilepton rates are summarized with various pion chemical potentials. Due to the mixing, the low invariant mass dilepton production is enhanced while the $\rho$-peak around 0.78 GeV is reduced indicating the partial restoration of chiral symmetry.

\subsection{Electric Conductivity}

\begin{figure}
\includegraphics[scale=.6]{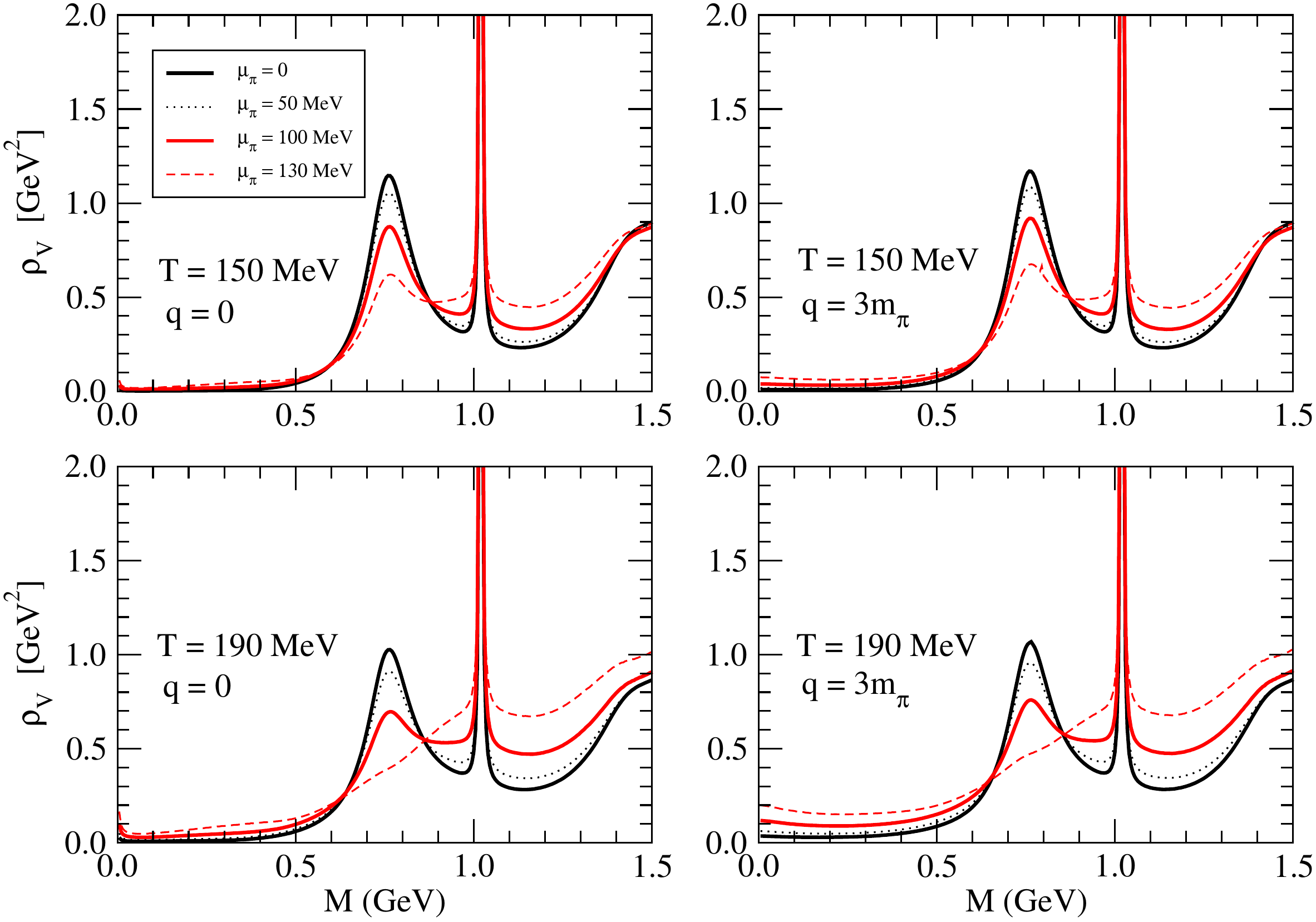}
\caption{Spectral function $\rho_{V}$  of the hadronic gas for $T=150$ \& 190 MeV with various $|\vec q|$ and $\mu_\pi$}
\label{fig:rhomu}
\end{figure}

\begin{figure}
\includegraphics[scale=.6]{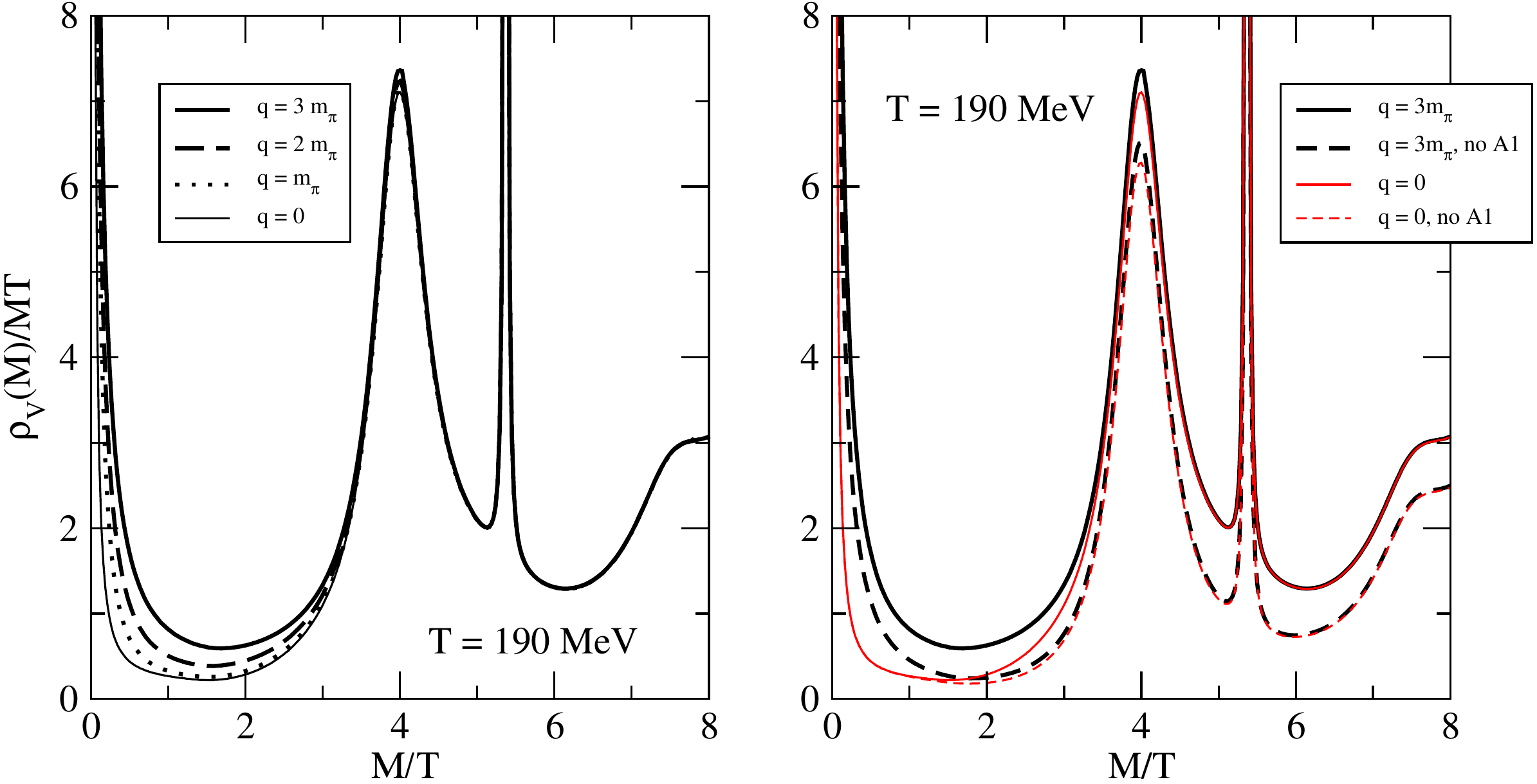}
\caption{$\rho_V/MT$ of the hadronic gas at $T=190$ MeV and $\mu_\pi=0$. The left panel shows 
the $|\vec q|$ dependence and the right panel shows the contribution of the A1 meson which is included in ${\rm Im}{\bf \Pi}_A$.}
\label{fig_rho_hadron}
\end{figure}

To assess the electric conductivity from the hadronic gas we can use linear response and the Kubo-like formula for the spectral function
\beqa
\rho_{V}(M,\vec q)= - \frac{2}{{\tilde {\bf e}^2}}\,{\rm Im}{\bf W}^R(M, \vec q)
\eeqa
where the sum of the squared flavor charge ratios ${\tilde {\bf e}}^2 \equiv \sum_f {\tilde e}_f^2$ and $\rho_V = - \rho_{00} + \rho_{ii}$\cite{Bur12}. 
In the $\vec{q} = 0$ limit 
\beqa
\rho_{ii} (M,\vec{0}) = \rho_V (M, \vec{0})
\eeqa
because the time-like component $\rho_{00}(M,\vec{0})$ vanishes due to  current conservation.
In Fig.~\ref{fig:rhomu} we show $\rho_{V}$ including terms upto $\kappa^2$ order for different values of $T, |\vec q|$, and $\mu_\pi$. As $\mu_\pi$ increases, one can clearly see the mixing between the vector and axial correlator. The contribution from $\phi$ remains largely unaffected by the hadronic medium effects
due to the OZI suppression rule. 
In Fig.~\ref{fig_rho_hadron} we summarize $\rho_{V}/MT$ for various values of $|\vec q|$ at $T=$190 MeV. In the left panel, one can see that the $\rho_{V}$ is enhanced as the momentum $\vec{q}$ increases especially in the low invariant mass region. In the right panel, we plot the same quantity with and without the A1 meson. In the region of $M/T = 1\sim 3$, the mixing between the vector and axial correlators are significant and the contribution of the A1 meson is very important.

\begin{figure}
\includegraphics[scale=.6]{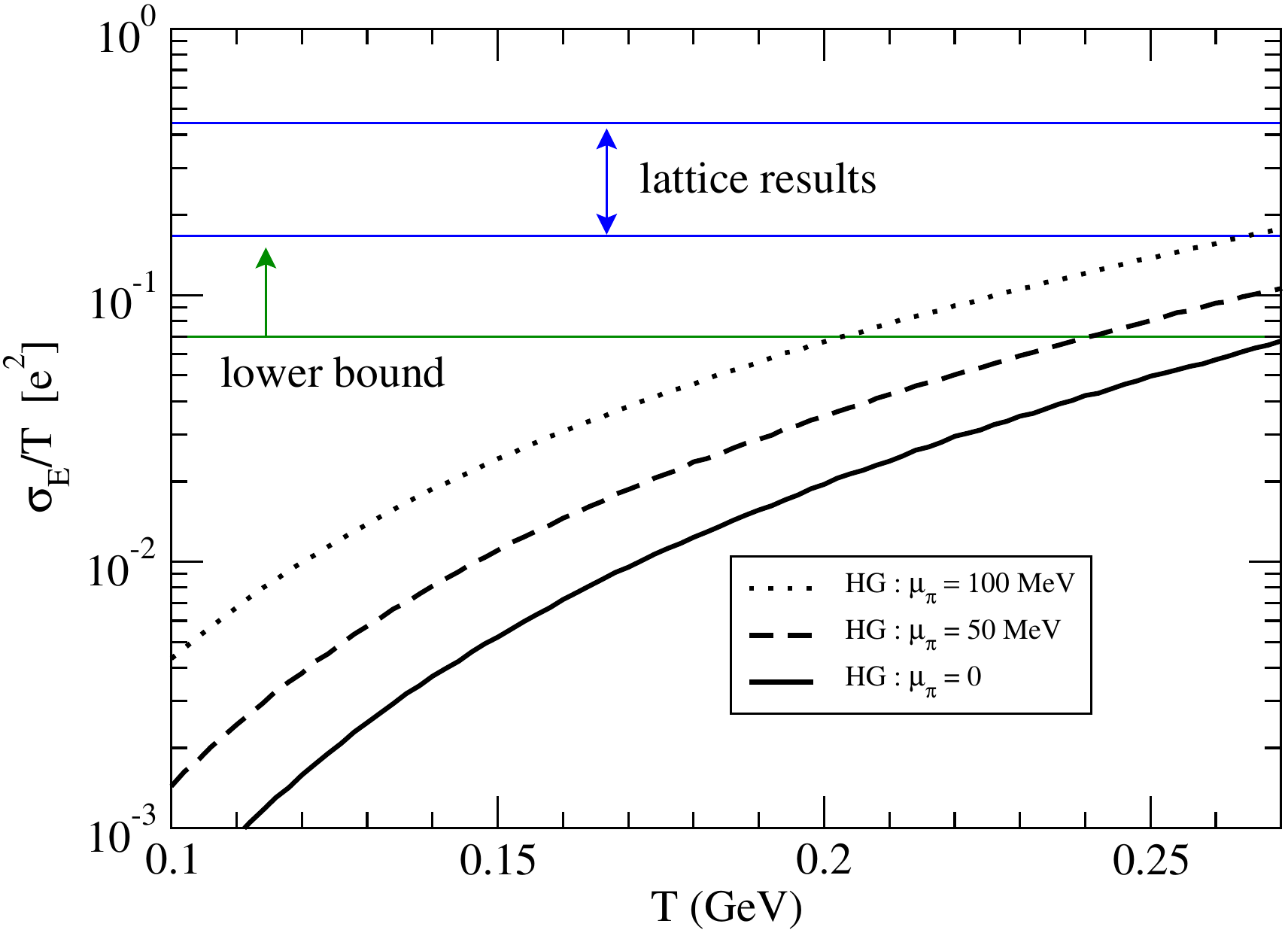}
\caption{$\sigma_E/T$ for the hadronic gas. 
The blue lines indicate the range of lattice results for 2 flavors~\cite{Din11,Kac13b}  and the green line indicates the lower bound \cite{Bur12}.
 }
\label{fig:econ}
\end{figure}

The electric conductivity  in unit of $e^2$ can be defined in the limit of $|\vec{q}|/M \rightarrow 0$ and $M\rightarrow 0$  
as
\beqa
\sigma_E= {\lim_{M\to 0}}\, \frac {{\tilde {\bf e}}^2{\rho_{ii}(M,\vec{0})}}{6M} = \lim_{M\to 0} \frac{-{\rm Im} {\bf W}^R(M,\vec{0})}{3M} 
= \lim_{M\to 0} \frac{-{\rm Im} {\bf W}^F(M,\vec{0})}{6T} . 
\label{eq:sigma}
\eeqa
One can easily confirm that there is no contribution to $\sigma_E$ from  ${\bf W}_\pi^F$   
because ${\rm Im} {\bf \Pi}_A(m_\pi^2) =0$.
In Fig.~\ref{fig_rho_hadron}, from the curves with $|\vec q|=0$, $\rho_{V}/MT$ increases very rapidly as we decrease $M$.    This behavior is caused by the pole of the retarded pion propagator in  ${\bf W}_{\pi\pi}^F$ in the region $\epsilon \ll M$. 
In order to separate the finite contribution from the hadronic gas, one can take the limit of $M/\epsilon \rightarrow 0$ for ${\rm Re}\Delta_R(k+q)$,
\be
\lim_{M/\epsilon \rightarrow 0} {\rm Re} (k+q) = \lim_{M/\epsilon \rightarrow 0} \frac{M^2+2 M E}{(M^2+2ME)^2 +\epsilon^4} 
\rightarrow 0.
\ee
In this limit, one can obtain a simple finite expression for the electric conductivity  to order $\kappa^2$,
\beqa
\frac{\sigma_E}{T}\approx \frac{(N_f^2-1)}{2T^2}\,\sum_{s=\pm}\,
\int \frac{d\pi_1}{f_\pi^2}\,\frac{d\pi_2}{f_\pi^2}\,(k_1+sk_2)^2\,{\rm Im}\Pi_V\left( (k_1+sk_2)^2 \right).
\label{SIGMAE}
\eeqa 
In Fig.~\ref{fig:econ} the electric conductivities from a hadronic gas are compared with recent lattice results~\cite{Din11,Kac13b} and the lower bound \cite{Bur12} which are discussed in Sec.~\ref{sec:econd-qgp}.
The $T$ and $\mu_\pi$ dependence of the hadronic gas is mainly caused by the pion distribution function. The hadron contribution to the electric conductivity is about an order of magnitude
smaller than the reported lattice results.

For completeness, we note that to one-loop in ChPT the vector spectral function in Eq.~(\ref{SIGMAE}) can be explicitly
assessed. The result for the electric conductivity is
\beqa
\frac{\sigma_E}{T}\approx \frac{(N_f^2-1)}{96\pi\,T^2}\,\sum_{s=\pm}\,
\int \frac{d\pi_1}{f_\pi^2}\,\frac{d\pi_2}{f_\pi^2}\Theta\left((k_1+sk_2)^2-4m_\pi^2\right)\,(k_1+sk_2)^2\,
\left(1-\frac{4m_\pi^2}{(k_1+sk_2)^2}\right)^{3/2}
\label{SIGMALOOP}
\eeqa 
which vanishes in the chiral limit as 
\beqa
\frac{\sigma_E}{T}\approx \frac{(N_f^2-1)T^4}{96\pi\,f_\pi^4}\,{\bf f}\left(\frac{m_\pi}{T}\right) =
\frac{(N_f^2-1)}{24\pi}\frac{\kappa^2m_\pi^2}{T^2}+{\cal O}\left(\frac{m_\pi^3}{T^3}\right)\; .
\eeqa
In the low temperature limit we have
\beqa
\frac{\sigma_E}{T}\approx \frac{(N_f^2-1)m_\pi^6}{96\pi T^2\,f_\pi^4}\,{\bf g}\left(\frac{T}{m_\pi}\right)
\eeqa
which is seen to vanish exponentially with the temperature  since ${\bf g}(T/m_\pi) \propto e^{-2m_\pi/T}$.  
 
\subsection{Quark number susceptibility}

The electric conductivity in unit of $e^2$ can be tied with the flavour diffusion constant $D_f$ through the identity
\cite{Bur12}
\be
\sigma_E=  \chi_f\left[\left(\sum_{f=1}^{N_f}{\tilde{e}}_f\right)^2\,{\bf D}_f^{\rm S}+
\left(\sum_{f=1}^{N_f}{\tilde{e}}_f^2\right)\,{\bf D}_f^{\rm NS}\right]
\ee
with ${\bf D}^{\rm S, NS}$ the singlet ($S$) and non-singlet $(NS)$ flavour diffusion constants and 
$\chi_f$ the flavour susceptibility
\be
\chi_f= \frac 1{TV_3}\,\left<{\bf Q}_f^2\right>
\ee
defined in terms of the conserved flavour charge
\be
{\bf Q}_f=\int\,d\vec{x}\,J_f^0(0,\vec{x}).
\ee
Note that  the singlet susceptibility vanishes for 3 flavours.

In the hadronic gas, the flavor susceptibility is better sought in terms of the fluctuations in the baryon number, isospin, and hyper-charge density through the linear transformation  
\be
\left(\begin{array}{c} {\bf Q}_u \\ {\bf Q}_d \\ {\bf Q}_s \end{array}\right) =
\left(\begin{array}{c c c}
1 & \;\;\; 1 \;\;\; & \frac{1}{2} \\
1 & -1 & \frac{1}{2} \\
1 & 0 & - 1 
\end{array}
\right) 
\left(\begin{array}{c} {\bf Q^B} \\ {\bf Q^I} \\ {\bf Q^Y} \end{array}\right) 
\ee
where 
\beqa
{\bf Q^B}  & = &  \int d\vec{x} \; q^\dagger \frac{{\bf 1}}{3}q
= \int d\vec{x} \; \frac{1}{3} \left(u^\dagger u + d^\dagger d + s^\dagger s \right) \nonumber\\
{\bf Q^I} &=& \int d\vec{x}\; q^\dagger \frac{\lambda^3}{2} q 
= \int d\vec{x}\; \frac{1}{2} \left(u^\dagger u  -d^\dagger d   \right) \nonumber\\
{\bf Q^Y} &=& \int d\vec{x}\; q^\dagger \frac{\lambda^8}{\sqrt{3}} q 
= \int d\vec{x}\; \frac{1}{3} \left(u^\dagger u + d^\dagger d -2  s^\dagger s \right).
\eeqa
Here ${\bf Q^B}$, ${\bf Q^I}$ and ${\bf Q^Y}$ correspond to the baryon number, isospin and hyper-charge operators, respectively.

In the pionic gas which we are considering in this work, the flavor susceptibility becomes flavour-dependent because the SU(3) symmetry is partially broken due to the explicit mass differences in the meson octet,
\beqa
\left( \begin{array}{c} \chi_u \\ \chi_d \\ \chi_s \end{array} \right)
= \frac{1}{TV_3}
\left(\begin{array}{c c c}
1 & \;\;\; 1 \;\;\; & \frac{1}{4} \\
1 &  1 & \frac{1}{4} \\
1 &  0 & 1 
\end{array}
\right) 
\left(\begin{array}{c} \langle ({\bf Q^B})^2 \rangle \\ \langle ({\bf Q^I})^2 \rangle \\ 
\langle ({\bf Q^Y})^2 \rangle \end{array}\right) 
\eeqa
where $\langle ({\bf Q^B})^2\rangle = \langle( {\bf Q^Y})^2 \rangle=0$ and $\chi_s=0$ for the pionic gas.

Using the pion density expansion we have
\be
\langle ({\bf Q^I})^2\rangle=\langle ({\bf Q^I})^2\rangle_\pi + \langle ({\bf Q^I})^2\rangle_{\pi\pi} +...
\ee
with
\be
\langle ({\bf Q^I})^2\rangle_\pi=\int d\pi \,\langle \pi^a(k)| ({\bf Q^I}) ^2|\pi^a(k)\rangle=
{{\bf I}_\pi^2}{V_3} N_\pi \int \frac{d^3k}{(2\pi)^3} n(E-\mu_\pi)
\ee
and 
\begin{eqnarray}
\langle ({\bf Q^I})^2\rangle_{\pi\pi}=&&+\frac 1{2!}\int d\pi^a(k_1) d\pi^b(k_2) \,
\left[\left<\pi^a(k_1)| ({\bf Q^I})^2|\pi^b(k_2)\right>\left<\pi^b(k_2)|\pi^a(k_1)\right>+  (a, k_1
\leftrightarrow b, k_2) \right]\nonumber\\
&&+\frac 1{2!}\int d\pi^a(k_1) d\pi^b(k_2) \,\,
{\rm Im}\left<\pi^a(k_1)\pi^b(k_2)|\left({\bf S}-{\bf 1}\right) ({\bf Q^I})^2|\pi^a(k_1)\pi^b(k_2)\right>
\end{eqnarray}
where ${\bf I}_\pi^2 = 2$, $N_\pi=3$,
$\langle \pi^b(k_2) | \pi^a(k_1) \rangle = \delta^{ab}(2\pi)^3 \; 2 E(k_1) \delta^3 (k_2 - k_1)$ 
and $ (2\pi)^3 \delta^3(\vec 0) = V_3 $.

\begin{figure}
\includegraphics[scale=.6]{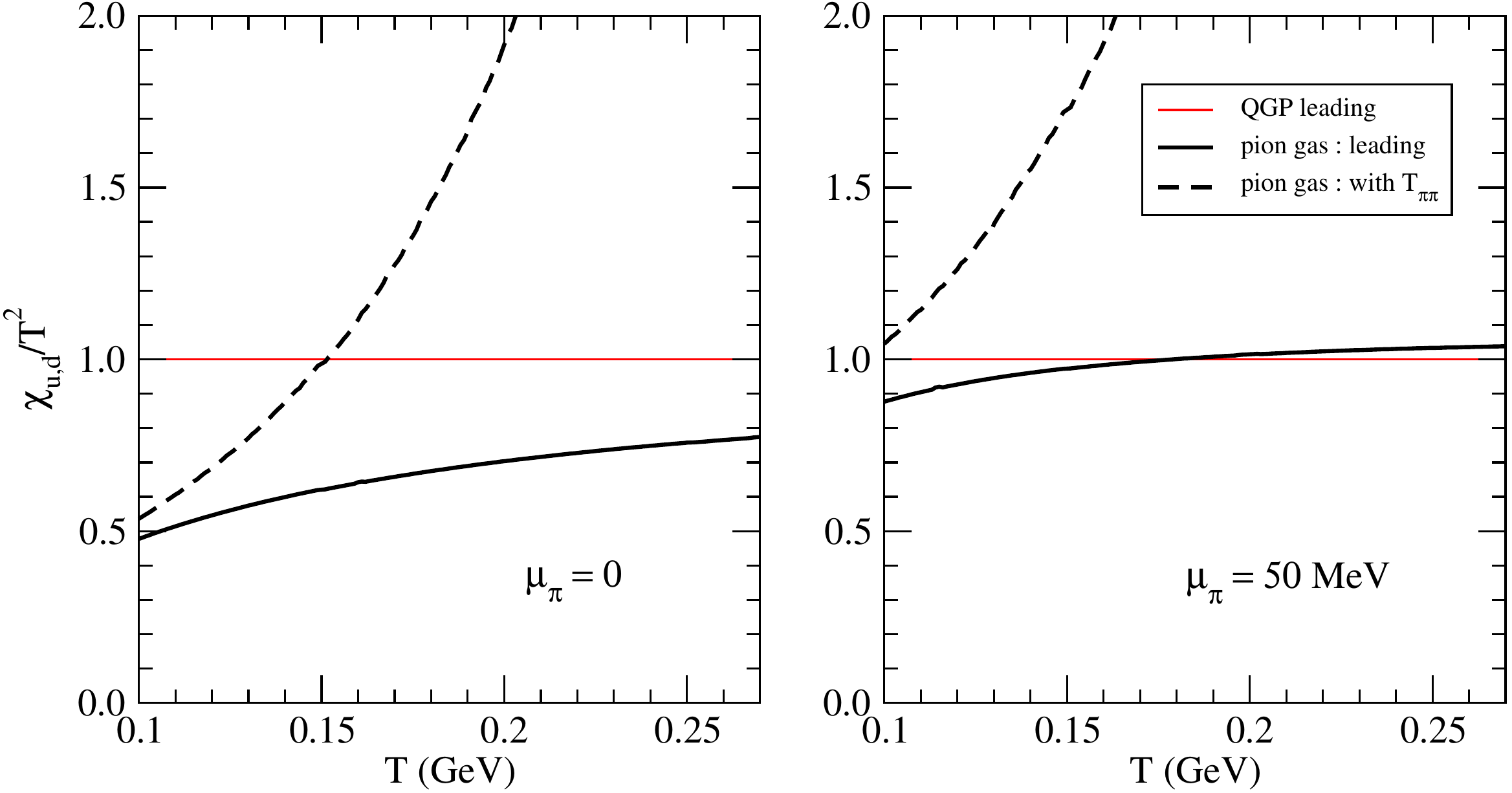}
\caption{Flavor susceptibilities of the pionic gas. Red thin solid line corresponds to the leading QGP result $\chi_{u,d}/T^2 = N_c/3$. The black thick solid line corresponds to the leading contribution with $\int d^3k\; n(1+n)$ and the black thick dashed line corresponds to the   susceptibilities with  ${\cal T}_{\pi\pi}$ as in Appendix~B. }
\label{fig:suscep}
\end{figure}

The first exchange but disconnected contribution is shown separately. The connected contribution 
involves  the full S-matrix after using  (\ref{XX1}). The result is the on-shell and forward $\pi\pi$
scattering amplitude ${\cal T}_{\pi\pi}$. The result is
\begin{eqnarray}
\left< ({\bf Q^I})^2\right>_{\pi\pi} & = & \frac{2{\bf I}_\pi^2}{2!}
{V_3} N_\pi \int \frac{d^3k}{(2\pi)^3} \left[n(E-\mu_\pi)\right]^2 \nonumber\\
&&+\frac {2{\bf I}_\pi^2}{2!}\int d\pi^a(k_1) d\pi^b(k_2) \,
(2\pi)^4\delta^4(k_1+k_2-(k_1+k_2))\,{\rm Re}{\cal T}_{\pi\pi}^{ab,ab}(k_1,k_2) .
\end{eqnarray}
where $(2\pi)^4 \delta^4(0)\equiv V_3 / T  $. Thus 
\begin{eqnarray}
\chi_{u,d} & = & \frac{1}{T V_3} \left< ({\bf Q^I})^2\right>  \nonumber\\
& \approx &
{\bf I}_\pi^2   
\left[\frac{N_\pi}{T} \int \frac{d^3k}{(2\pi)^3} n \left(1+n \right) 
+    \frac{1}{T^2} \int \frac{d^3 k_1}{(2\pi)^3}\frac{n_1}{2 E_1} \frac{d^3 k_2}{(2\pi)^3}\frac{n_2}{2 E_2}  \; {\rm Re}{\cal T}_{\pi\pi} (s,t,u)
\right]
\end{eqnarray}
with the Mandelstam variables $s=(k_1+k_2)^2$, $t=(k_1-k_2)^2$, $u=0$.
To leading order in ChPT the $\pi\pi$ scattering amplitude is given by the Weinberg term. Specifically,

\begin{eqnarray}
\chi_{u,d} \approx {\bf I}_\pi^2   
\left[\frac{N_\pi}{T} \int \frac{d^3k}{(2\pi)^3} n \left(1+n \right) 
-  \kappa^2 N_\pi(N_\pi-2) \frac{m_\pi^2f_\pi^2}{T^2} \,
\right]
\end{eqnarray}
where the tree level $\pi\pi$ contribution is seen to be negative and vanishing in the chiral limit.
The full result for the second order correction using the chirally reduced forward $\pi\pi$-scattering amplitude is given in the Appendix~B in terms of the pion scalar and vector form factors and vacuum correlators \cite{Yam96a,Yam96b}. 
In Fig.~\ref{fig:suscep}, the flavor susceptibilities of the pionic gas are summarized. At low temperature the leading contribution dominates compared to the ${\cal T}_{\pi\pi}$ contribution. However, as the temperature increases, the ${\cal T}_{\pi\pi}$ contribution dominates due to the extra $T^2$ dependence compared to the leading contribution. In this high temperature region, the perturbative description of the pionic gas is not valid. In Fig.~\ref{fig:suscep} the leading QGP contribution is given by the red thin lines. Higher order corrections to the sQGP susceptibility are discussed later in Sec.~\ref{sec:qgpdif}.

\section{Electromagnetic Radiation from a Strongly Interacting Quark-Gluon Plasma}

\label{sec:sQGP}

\subsection{Non-Perturbative Thermal Condensates}

There has been great progress in the calculation of the perturbative photon emission rates in a weakly coupled QCD plasma at 
asymptotic temperatures~\cite{Arn01}. 
The leading contribution to the photon rates comes from two-loop diagrams corresponding to the process $q+\overline{q}\to\gamma + g$ and compton $g+q(\overline{q}) \to q(\overline{q}) + \gamma$ processes.  However these rates are plagued with collinear singularities. Instead, a complete leading order photon emission requires the inclusion of collinear bremsstralung and inelastic pair annihilations  and their subsequent suppression through the LPM effect \cite{Arn01}. The extension of these
calculations to the dilepton rates at asymptotic temperatures is not available.

At current collider energies the QCD plasma is strongly coupled or sQGP. The perturbative calculations are at best
suggestive and a more non-perturbative framework for time-like processes is needed to separate the hard partonic
physics which is perturbative from the soft partonic physics which is not. A useful framework for this approach is the
one advocated long ago by Hansson and one of us~\cite{Han87} whereby the vacuum OPE expansion for
current-current correlators is re-ordered at high temperature to account for the soft thermal gluon corrections 
through pertinent electric and magnetic condensates much in the spirit of the QCD-sum-rules in the non-perturbative
vacuum. Its application to thermal dileptons was already used in~\cite{Lee99}.

\begin{figure}
\includegraphics[scale=.6]{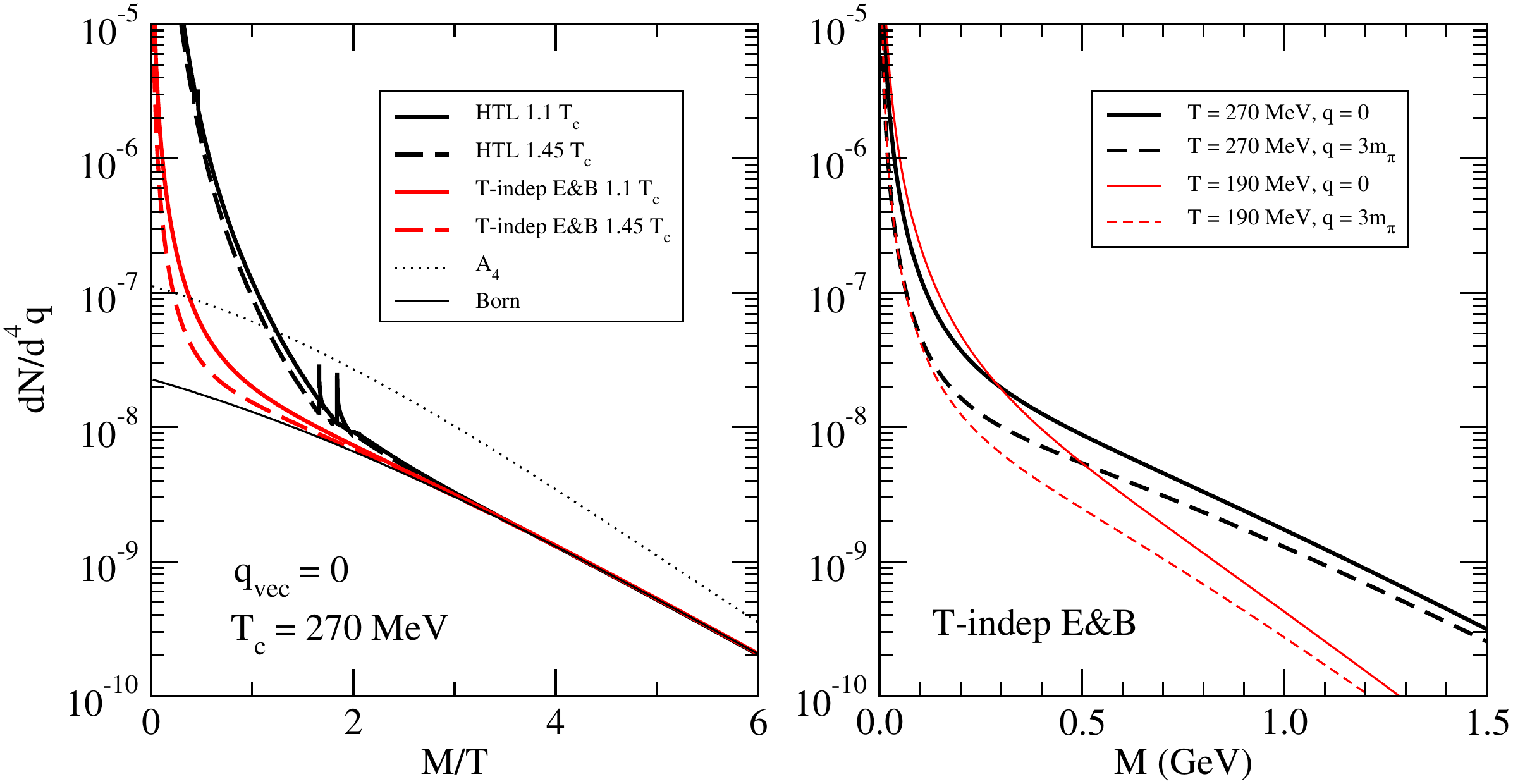}
\caption{Thermal dileptons; left panel: sQGP and HTL, right panel: $T\& |\vec q|$ dependence of  the sQGP. 
$T$-independent $\langle   B^2 \rangle$ and $\langle  E^2 \rangle$ in Eq.~(\ref{eq:g2}) are used for the sQGP.}
\label{fig_dNdq}
\end{figure}

The approach works as follows: The leading order contribution to the retarted current-current
correlator, Eq.~(\ref{eer}), is the "Born" $q\bar q$ annihilation term,
\beqa
{\rm Im}\,{\bf W}_0^R(q)=\frac {N_c {\tilde {\bf e}}^2}{4\pi} \,q^2\,\left[1+\frac{2T}{|\vec{q}|}\,{\rm ln}\left(\frac{n_+}{n_-}\right)\right]
\label{PERT}
\eeqa
where $N_c$ is the number of colors and $n_\pm$ the quark occupation numbers
\beqa
n_\pm =\frac 1{e^{(q_0\pm |\vec{q}|)/2T}+1} \, .
\eeqa
Note that this contribution vanishes at the photon point, $q^2=0$, due to energy momentum conservation \cite{Lee99}.
 The sQGP around the critical temperature is expected to display non-perturbative effects in the form of soft gluons, which can be characterized by thermal condensates of  gauge-invariant operators of leading mass dimensions
such as $\left<A_4^2\right>$, $\left<E^2\right>$
and $\left<B^2\right>$.  Their contributions to the dilepton emissivities in leading order are~\cite{Han87,Lee99}
\beqa
{\rm Im}\,{\bf W}_2^R(q)=\frac{N_c {\tilde {\bf e}}^2}{4\pi} q^2
 \left<\frac {\alpha_s}\pi A_4^2\right>\left(\frac{4\pi^2}{T|\vec{q}|}\right)\left(n_+(1-n_+)-n_-(1-n_-)\right) 
\label{AA}
\eeqa
and
\beqa
{\rm Im}\,{\bf W}_4^R(q)=\frac{N_c {\tilde {\bf e}}^2}{4\pi} 
\left[-\frac 16\left<\frac {\alpha_s}\pi E^2\right>+\frac 13\left<\frac {\alpha_s}\pi B^2\right>\right]
\left(\frac{4\pi^2}{T|\vec{q}|}\right)\left(n_+(1-n_+)-n_-(1-n_-)\right) .
\label{GG}
\eeqa
Across the phase transition temperature $T_c$ which is first order for pure gluo-dynamics, the electric and magnetic
condensates fall by about half their value in the QCD vacuum in the temperature range (1$-$3)$T_c$, and remain 
about constant in this range \cite{Ada91}. Thus for $T_c<T<3T_c$ in Euclidean signature this translates to 
\beqa
\left< {\alpha_s} B^2\right>\approx \left<{\alpha_s} E^2\right>\approx 
\frac 12 \times  \frac 14 \left<{\alpha_s} G^2\right>_0 
\label{eq:g2}
\eeqa
in terms of the vacuum gluon condensate \cite{Ada91}. We use the updated value of the gluon condensate
$\langle \alpha_s G^2\rangle_0 = 0.068$~GeV$^4$ \cite{Nar09}. 
In Fig.~\ref{fig_dNdq}, the dilepton rates from the sQGP are summarized for various temperatures and momenta $q=|\vec q|$. 
In order to check the contribution from $\langle A_4^2 \rangle$, we used $\langle \frac{\alpha_s}{\pi} A_4^2 \rangle/T^2 \approx 0.4$ for the plot \cite{Lee99}. The presence of $\left<A_4^2\right>$ appears to be ruled out by a comparison to the recent lattice data~\cite{Kac13}.  In the left panel, for the comparison, we also plot the contribution form the HTL (hard thermal loop) \cite{Bra90}. One can see that the enhancement in the low mass region mainly comes from the $\langle E^2\rangle$ and $\langle B^2 \rangle$ contributions even though they are smaller than the HTL results. In Braaten et al. \cite{Bra90}, power counting is taken into account even for the Fermi-Dirac distribution function, which is valid in the soft energy region. However, in this work, we kept the full expression of the Fermi-Dirac distribution function in the HTL calculation in order to compare with other results. In the right panel of Fig.~\ref{fig_dNdq}, the temperature and momentum dependence of the sQGP rate are summarized. By comparing results for $T=190$ MeV in Figs.~\ref{fig_dNdq_had} and \ref{fig_dNdq}, one can see that the hadronic contributions are significantly higher than the sQGP contribution in the low mass region as the chemical potential increases.

\begin{figure}
\includegraphics[scale=.6]{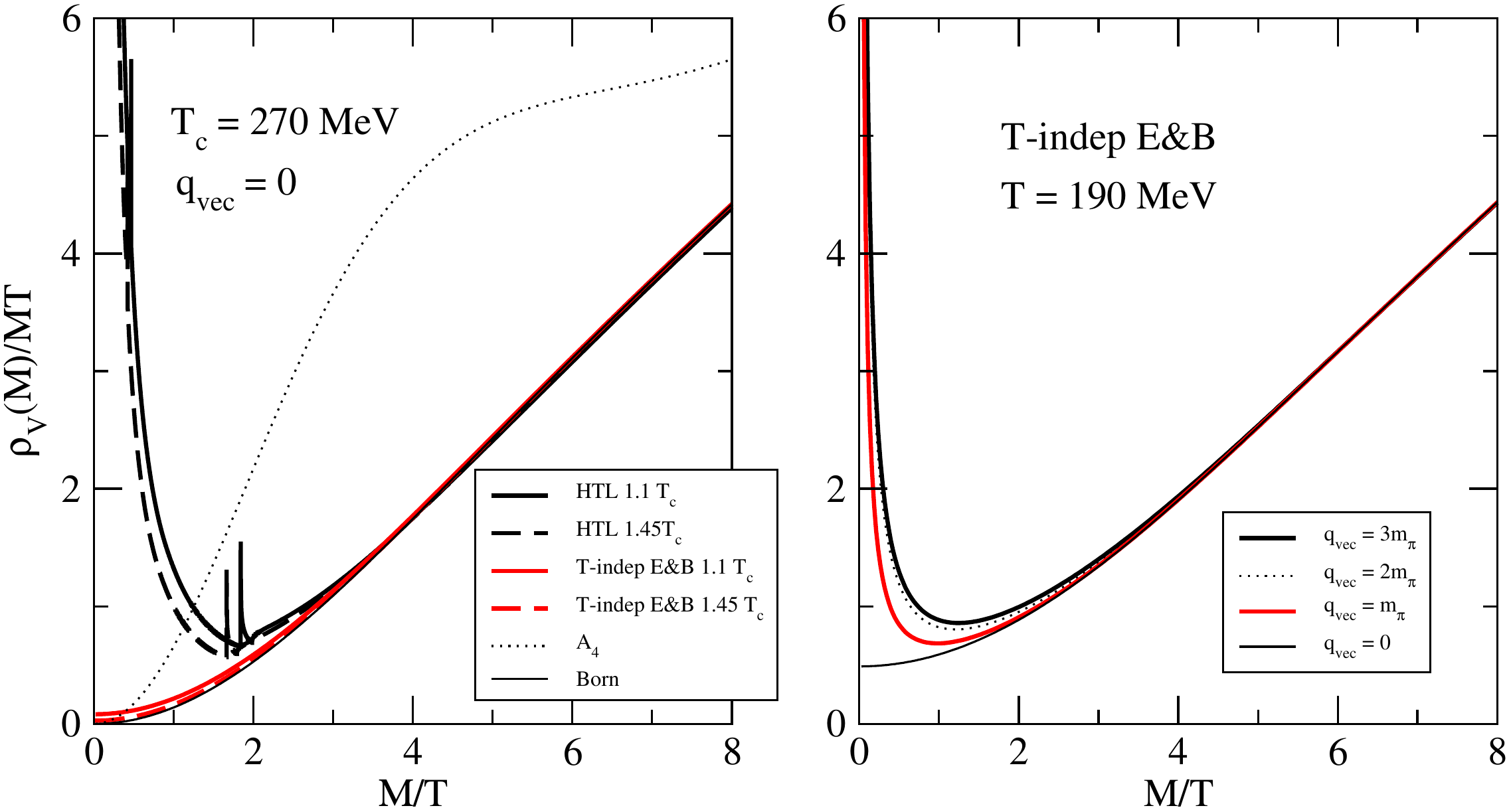}
\caption{Vector spectral density: left panel; comparison between sQGP and HTL results, right panel: $|\vec q|$ dependence of sQGP. $T$-independent $\langle   B^2 \rangle$ and $\langle  E^2 \rangle$ in Eq.~(\ref{eq:g2}) are used for sQGP. }
\label{fig_rho}
\end{figure}

In Fig.~\ref{fig_rho} we plot the vector spectral densities, which can be compared with the results from the
hadronic gas summarized in Fig.~\ref{fig_rho_hadron}.
In the left panel of Fig.~\ref{fig_rho} we compare the results at two high temperatures of $1.1\,{\rm T_c}$ and $1.45\,{\rm T_c}$ with the critical temperature for quenched calculation $T_c=$ 270 MeV  \cite{Din11}.  
The leading Born contribution is compared to the contribution including the soft gluon condensates as well as the hard thermal loops
\cite{Bra90,Wel89}. In the right panel   the same spectral densities are shown for different momenta $\vec{q}\neq 0$ at $T=190$ MeV. 
With finite thermal condensate $\langle E^2\rangle$ and $\langle B^2\rangle$ contribution, the $\rho_{V}/MT$ increases as the momentum increases for any given $M$, especially in the low mass region the enhancement is significant. A comparison with recent lattice results confirms the important of the thermal condensate in the sQGP\cite{Kac13}.

\subsection{Electric Conductivity}
\label{sec:econd-qgp}

The electric conductivity $\sigma_E$ at high temperature plays an important role in recent developments related 
to the chiral magnetic effects in the early stage of the sQGP.   Our condensate corrections to the Euclidean spectral function allow us to make an estimate of $\sigma_E$ across the transition region by tying it to the spectral function in the zero mass limit as in Eq.~(\ref{eq:sigma}). The only drawback  is that the re-organized OPE expansion at high temperature~\cite{Han87,Lee99}  is an expansion
in ${\bf M}^2/|\vec{q}|^2<1$, with ${\bf M}$ the soft scale in the matrix element which is typically the magnetic scale. The extrapolation of the leading operator corrections to $|\vec{q}|\rightarrow 0$ while finite calls for corrections of order 1 
from the higher operator insertions. This notwithstanding, an estimate of the electric conductivity is set by the leading 
dimension 4 operators at high temperature
\beqa
\sigma_E \approx   
 \frac {\pi N_c {\tilde {\bf e}}^2}{48T^3} 
\left(-\frac 16\left<\frac {\alpha_s}\pi E^2\right>+\frac 13\left<\frac {\alpha_s}\pi B^2\right>\right).
\eeqa
Lattice results show that $\sigma_E/T$ is weakly dependent on the temperature and the value lies  in the range
$0.3 < \sigma_E / {\tilde {\bf e}}^2 T < 0.8 $ \cite{Din11,Kac13b}.
Recent analysis with PHENIX data  gives slightly larger value $0.5 < \sigma_E/T < 1.1$ \cite{Yin13}.
The temperature dependence of $\sigma_E/T$ has been also reported in the literature~\cite{Ama13,Ste13,Cas13,Mar13}, in which $\sigma_E/T$ increases as the temperature increases above $T_c$.
Burnier \& Laine \cite{Bur12} got a  lower bound for the electric conductivity, or $\sigma_E/T \ge 0.07$, which is significantly smaller than previous  leading-order weak-coupling expansion results~\cite{Arn00,Arn03}.

\begin{figure}
\includegraphics[scale=.6]{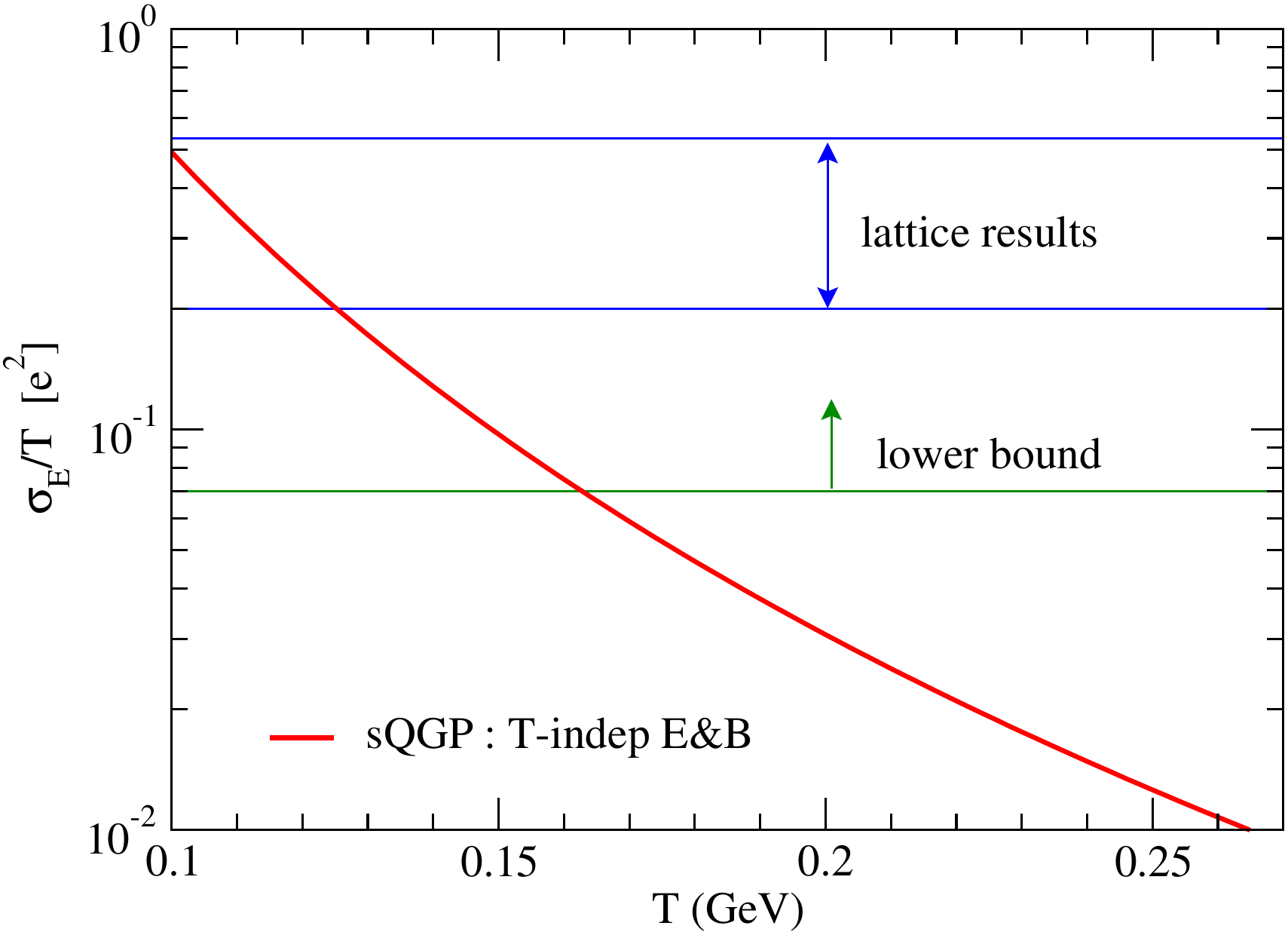}
\caption{$\sigma_E/T$ for sQGP. 
The blue lines indicate the range of lattice results for 3 flavors~\cite{Din11,Kac13b}  and the green line indicates the lower bound \cite{Bur12}.
 }
\label{fig:econ-qgp}
\end{figure}

\begin{figure}
\includegraphics[scale=.6]{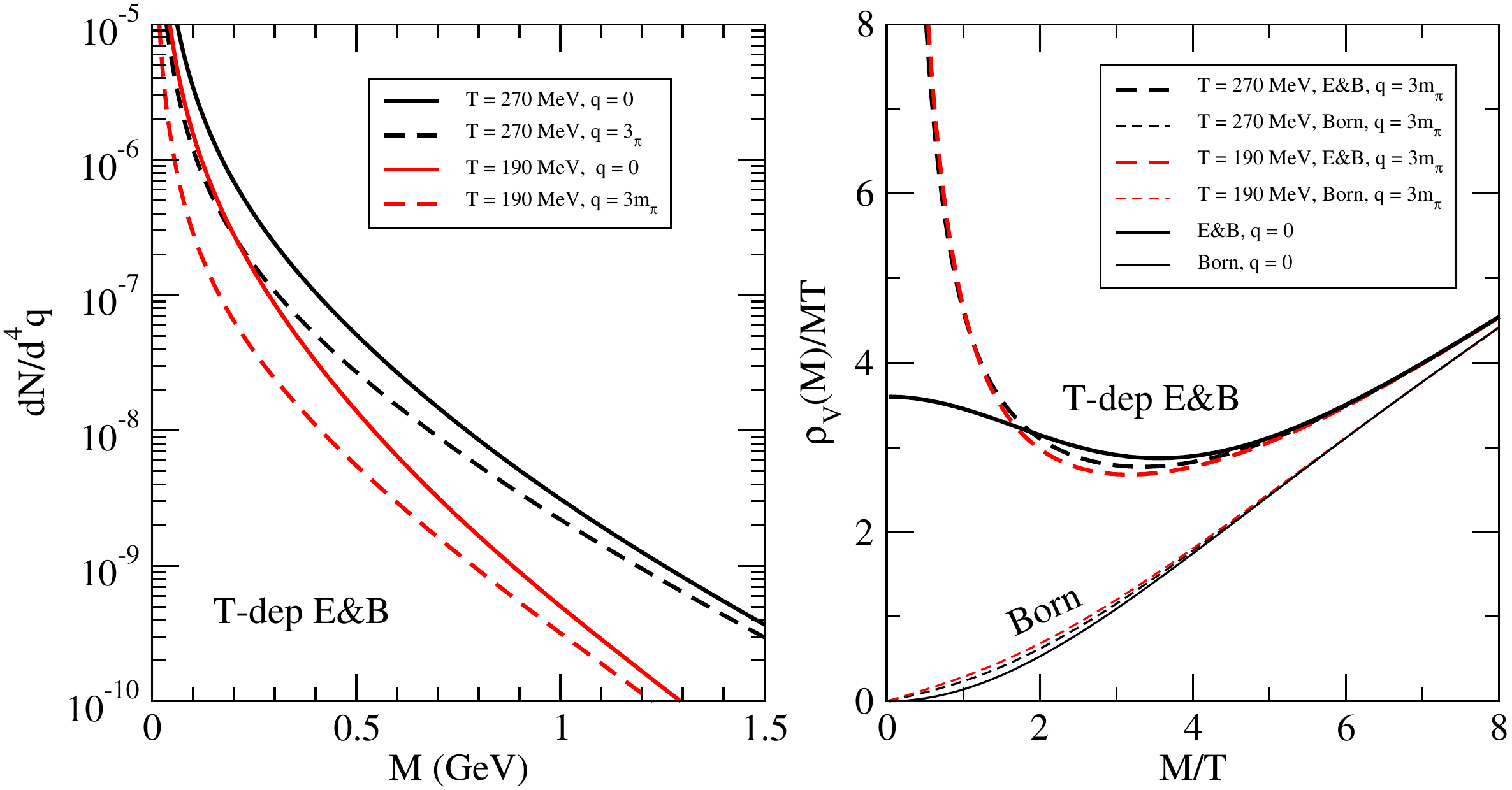}
\caption{Dilepton rates and spectral function from the sQGP with $T$-dependent $\langle  E^2 \rangle$ and $\langle   B^2 \rangle$ in
Eq.~(\ref{eq:Tdep-cond}).}
\label{fig:Tdep-cond}
\end{figure}

In Fig.~\ref{fig:econ-qgp}, we plot the electric conductivity for  the sQGP with constant $\langle B^2 \rangle$ and $\langle E^2\rangle$.  
Our sQGP results with constant $\langle B^2 \rangle$ and $\langle E^2\rangle$ are  much smaller than the lattice estimates~\cite{Din11,Kac13b}. At large temperatures the electric and magnetic condensates are $T$ dependent  with $\langle    B^2 \rangle\approx \langle  E^2 \rangle\approx (b \pi^2/20)\times T^4$ and $b\approx 1-1.2$~\cite{Ele93}. On the other hand, a fit to the currently
reported  lattice conductivities suggest 
\beqa
\langle \alpha_s E^2 \rangle \approx \langle \alpha_s B^2 \rangle \approx \frac{288}{N_c } 
\left\langle \frac{\sigma_E}{{\tilde {\bf e}}^2 T} \right\rangle  T^4 \approx 48\, T^4
\label{eq:Tdep-cond}
\eeqa
with $\langle \sigma_E/{\tilde {\bf e}}^2T\rangle \sim 0.5$ at about the mean value of the lattice  results  \cite{Din11,Kac13b}. 
In Fig.~\ref{fig:Tdep-cond}, we plot the dilepton rate and spectral density with $T$-dependent condensates.  In comparison with Figs.~\ref{fig_dNdq} and \ref{fig_rho} one can see the significant enhancement in the low mass region. The Born term dominates in the high mass region and the results are rather insensitive to the details of the thermal condensates.

\subsection{Flavor Diffusion Constant}
\label{sec:qgpdif}

The partonic flavor susceptibility can be sought along the same arguments as those developed for
the hadronic parts, using the QCD Hamiltonian at high temperature. Indeed, 
\be
\left<{\bf Q}_f^2\right>=\left<{\bf Q}_f^2\right>_q + \left<{\bf Q}_f^2\right>_{qq} +...
\ee
with 
\be
\left<{\bf Q}_f^2\right>_q=\int dq(k) \,\left<q^{ai}(k)|{\bf Q}_f^2|q^{ai}(k)\right>=
4N_c{V_3}\int \frac{d^3k}{(2\pi)^3} n_F(E)
\ee
and
\begin{eqnarray}
\left<{\bf Q}_f^2\right>_{qq}=&&-\frac {1}{2!}\int dq(k_1) dq(k_2) \,
\left(\left<q^{ai}(k_1)|{\bf Q}_f^2|q^{bj}(k_2)\right>\left<q^{bj}(k_2)|q^{ai}(k_1)\right>+ a, i, k_1
\leftrightarrow b, j,  k_2 \right)\nonumber\\
&&+\frac {1}{2!}\int dq(k_1) dq(k_2) \,\,
{\rm Im}\left<q^{ai}(k_1)q^{bj}(k_2)|\left({\bf S}-{\bf 1}\right){\bf Q}_f^2| q^{ai}(k_1)q^{bj}(k_2)\right>\; .
\end{eqnarray}
The index $a$ is for flavor and the index $i$ is short for color, spin, particle and anti-particle.
The integrals count the number of massless (scalar) fermions in phase space
\be
\int dq(k)=\int \frac{d^3k}{(2\pi )^3}\, \frac {n_F(E)}{2E}.
\ee
In the the disconnected matrix element the minus sign is from the antisymmetric switch of the
quarks. The connected contribution is the forward quark-quark scattering amplitude
${\cal T}_{qq}$. Thus
\begin{eqnarray}
\left<{\bf Q}_f^2\right>_{qq}=&&-4N_cV_3\int \frac{d^3k}{(2\pi)^3} n_F^2(E)
\nonumber\\
&&+\frac {2}{2!}\int dq(k_1) dq(k_2) \,\,
\,(2\pi)^4\,\delta(k_1+k_2-(k_1+k_2))\,{\rm Re}{\cal T}^{ai,bj}_{qq}(k_1,k_2)
\end{eqnarray}
so that
\begin{eqnarray}
\chi_f \approx && \frac{4N_c}{T} \int \frac{d^3k}{(2\pi)^3} n_F (1-n_F) 
+\frac {1}{T^2}\int dq (k_1) dq (k_2)  {\rm Re}{\cal T}^{ai,bj}_{qq}(k_1,k_2).
\end{eqnarray}
For massless quarks, the first term gives the leading QGP contribution $\chi_f = (N_c/3) T^2$.  In this work, instead of calculating the contributions of $T_{qq}$ explicitly, we compare our pionic gas results in Fig.~\ref{fig:suscep} with the recent lattice results \cite{Baz13}. Lattice results indicate that the quark susceptibilities drop by about $15 \sim 25$ \% compared to the Stefan-Boltzmann limit near the phase transition temperature. The leading contribution from the pionic gas is close to the lattice results. However, as noted earlier, the higher order corrections from ${\cal T}_{\pi\pi}$ become significant in the critical temperature region and the perturbative treatment is not valid.

Since ${\bf D}_f^{NS}/{\bf D}^S_f\sim N_c$ this makes the non-singlet contribution dominant
for $N_c=3$ assumed large. Thus, with $\chi_f=(N_c/3) T^2$,
\be
TD_f^{NS}\approx \frac{T \sigma_E}{{\tilde {\bf e}}^2 \chi_f} 
\approx
\frac{3}{N_c} \frac{\sigma_E}{T{\tilde {\bf e}}^2}\approx \frac 12
\ee
where in the last estimate we used the central value of the lattice estimate for the electric conductivity,
across the transition temperature. 
In the intermediate regime of temperatures $(1-3)\,T_c$ the light flavour quarks carry a thermal
mass of the order of the Matsubara mass $m_T\approx \pi T>T$ making the light flavors somehow 
heavy  in comparison to the typical thermal excitations. In the large $N_c$ limit and 
using the Einstein relation we can estimate  the drag $\eta_f$ on the light quarks
in the transition region \cite{Moo05,Cao13}
\be
\frac{\eta_f}T\approx \frac{1}{m_T{ \bf D}^{NS}_f}\approx \frac{N_cT}{3m_T}
\frac{ {\tilde {\bf e}}^2 T}{\sigma_E}\; .
\ee
If we use the central value of the lattice result for the   electric conductivity, then  $\eta_f/T\approx 2/\pi$
across the transition temperature. This drag quantifies the amount of Brownian motion for the light flavors in the sQGP.

\section{Conclusions}
\label{sec:con}

Our hadronic rates are based on the use of spectral functions. Unlike kinetic processes whereby each 
emission is associated with particular Feynman diagrams, our spectral analysis enforces
all the constraints of broken chiral symmetry, and through the spectral weights accounts for 
tails of resonances. It does not rely on any effective Lagrangian, and therefore does not
suffer the drawback of a strong interaction expansion and the ambiguities associated to
hadronic form factors. However, it is limited by a reorganization of the leptonic 
emissivities around the resonance gas model to leading order, with 
one- and two-pion final re-scattering in the initial states. Carrying out the expansion to three-pion
re-scattering in the initial state  is formidable.

We have shown that the mixing between the vector and axial correlators becomes more significant with increasing
pion chemical potentials indicating the partial restoration of chiral symmetry. This mixing enhances the dilepton  rate significantly at low invariant mass.   The evolved rates account well for the dilepton emissivities reported by the 
SPS (see \cite{Rap13} and reference there in).  Although the inclusion of baryons, should improve slightly the
fit, we are confident that our organization of the dilepton emissivities through the virial expansion works at collider
energies.

Since our photon rates fit reasonably well the low mass photon spectra at collider energies~\cite{Dus10}
we can use them to extract both the electric conductivity and the flavor susceptibility constant in the hadronic phase. 
We have found that the electric conductivity at $T\approx m_\pi$ is substantially smaller than the currently reported
lattice conductivities. While we have not included the contributions of order $\kappa^3$ and higher, we
believe that our chiral expansion provides a sound starting estimate based on the strictures of spontaneously
broken chiral symmetry. The flavor susceptibility in the correlated hadronic gas is reasonably close to the
reported lattice results at the transition temperature.

We have provided first principle estimates of the corrections to the electromagnetic emissivities in the partonic
phase and near the transition temperature using the high temperature QCD sum rule method~\cite{Han87,Lee99},
whereby the effects of soft gluons are retained in the form of gluonic matrix elements. A reasonable account of the
electric conductivities reported on the lattice at high temperature is reproduced with temperature dependent
condensates. 

The approach we have discussed can be extended to most  transport coefficients in QCD both below and
above the transition temperature.  It is well motivated by the structures of chiral symmetry below the transition
temperature, and by a reorganization of the OPE expansion at high temperature. The dual nature of the
interacting resonance gas model near the transition temperature with its high-temperature partonic description,
provides us with an interesting non-perturbative tool for computing the transport parameters of QCD matter near
equilibrium.

\section*{Acknowledgements}

We would like to thank K.~Dusling, S.~Jeon, R.~Rapp, and D.~Teaney for  helpful discussions.
The work of CHL was supported by the BAERI Nuclear R \& D program (M20808740002) of MEST/KOSEF and the Financial Supporting Project of Long-term Overseas Dispatch of PNU's Tenure-track Faculty, 2013.
The work of IZ was supported in part by US DOE grants DE-FG02-88ER40388
and DE-FG03-97ER4014.

\section*{Appendix A. Two-Pion Contribution ${\bf W}_{\pi\pi}^F$}

The two-pion contribution ${\bf W}^F_{\pi\pi}$ which is important both for the rate and the electric conductivity is more involved~\cite{Ste97,Dus09b}. We summarize the dominant contributions \cite{Dus10}
\begin{eqnarray}
\frac{1}{f_\pi^4} {\rm Im} {\bf W}^F_{\pi\pi}(q,k_1,k_2) 
&=& \frac{2}{f_\pi^2}\left[g_{\mu\nu}-(2k_1+q)_\mu k_{1\nu}\text{Re}\Delta_R(k_1+q)\right]\text{Im}\mathcal{T}_{\pi\gamma}^{\mu\nu}\left(q,k_2\right)\nn
&+& (q\to -q) + (k_1\to -k_1) + (q,k_1 \to -q,-k_1)\label{eq:2pig}\nn
&+& \frac{1}{f_\pi^2}k_1^\mu (2k_1+q)^\nu \text{Re}\Delta_R(k_1+q)\epsilon^{a3e}\epsilon^{e3g}\text{Im}\mathcal{B}^{ag}_{\mu\nu}(k_1,k_2)\nn
&-& \frac{1}{f_\pi^2}\left[g^{\mu\nu} - (k_1+q)^\mu(2k_1+q)^\nu\text{Re}\Delta_R(k_1+q)\right]\nn
&& \times \epsilon^{a3e}\epsilon^{a3f}\text{Im}\mathcal{B}_{\mu\nu}^{ef}(k_1+q,k_2)\nn
&+& \frac{1}{f_\pi^2}(k_1+q)^\mu(k_1+q)^\nu(2k_1+q)^2\left[\text{Re}\Delta_R(k_1+q)\right]^2\nn
& & \times \epsilon^{a3e}\epsilon^{a3f}\text{Im}\mathcal{B}_{\mu\nu}^{ef}(k_1+q,k_2) + (k_1\to-k_1).
\label{eq:2B}
\end{eqnarray}
The pion-spin averaged $\pi\gamma$ forward scattering amplitude ${\rm Im} {\cal T}_{\pi\gamma}$  is given as \cite{Dus09b}  
\beqa
{\rm Im} {\cal T}_{\pi\gamma}^{\mu\nu} (q,k)
&=& \frac{2}{3 f_\pi^2} (2k^\mu+q^\mu) (-q^2 k^\nu + k\cdot q\, q^\nu) {\rm Re}\Delta_R(k+q)
{\rm Im}{\bf \Pi}_V(q^2) \\
&+& (q \rightarrow -q) + (k \rightarrow -k) + (q,k \rightarrow -q,-k) \nn
&+& \frac{4}{3 f_\pi^2} (g^{\mu\nu} q^2 - q^\mu q^\nu) {\rm Im} {\bf \Pi}_V (q^2) \\
&-& \frac{2}{3 f_\pi^2} \left( g^{\mu\nu} (k+q)^2 - (k+q)^\mu (k+q)^\nu \right) {\rm Im} {\bf \Pi}_A\left((k+q)^2\right) \\
&+& (k \rightarrow -k ),
\eeqa
and the contribution  $\mathcal{B}$ reads  \cite{Dus10,Dus09b}
\begin{eqnarray}
\text{Im}\mathcal{B}_{\mu\nu}^{ef}(k_1,k_2)&=&\frac{2}{f_\pi^2}\delta^{ef}\left[ g_{\mu\nu}(k_1+k_2)^2-(k_1+k_2)_\mu (k_1+k_2)_\nu\right]\text{Im}\Pi_V\left( (k_1+k_2)^2 \right)\nn &+& (k_2\to-k_2)\nn
&-&\frac{4}{f_\pi^2}\delta^{ef}\left[ g_{\mu\nu}k_1^2-k_{1\mu} k_{1\nu}\right]\text{Im}\Pi_A\left( k_1^2 \right).
\label{eq:Bfinal}
\end{eqnarray}
All additional spectral contributions to ${\bf W}_{\pi\pi}^F$  are thoroughly discussed in
\cite{Ste97,Dus09b}. Their contribution to the photon and dilepton emissivities  in the low and intermediate
mass range is negligible.

\section*{Appendix B. $\pi\pi$ Scattering Amplitude}

Here we summarize the $\pi\pi$ scattering amplitudes which are relevant to the flavor susceptibility as \cite{Yam96a,Yam96b}
\beqa
{\cal T}_{\pi\pi} (s,t,u) & \equiv & \sum_{a=d,b=c} 
\left. {\cal T}_{\pi\pi} (p_2 d, k_2 b \leftarrow k_1 a, p_1 c) \right|_{p_2=k_1, p_1=k_2} \nonumber\\
& = &   {\cal T}_{\rm tree} (s,t,u) +  {\cal T}_{\rm vector} (s,t,u) +  {\cal T}_{\rm scalar} (s,t,u)
+  {\cal T}_{\rm rest} (s,t,u)
\eeqa 
with Mandelstam variables
\beqa
s &=& (k_1 + p_1)^2 = (k_2 + p_2)^2 \nonumber\\
t &=& (k_1- k_2)^2 = (p_1 - p_2 )^2 \nonumber\\
u &=& (k_1 -p_2)^2 = (p_1 - k_2)^2.
\eeqa
For the contribution of thermal pions to the flavor susceptibility, $\delta^{ad}\delta^4(k_1-p_2)$,
$\delta^{bc}\delta^4(p_1-k_2)$, 
and $u=0$ are implicitly considered and the identity $s+t+u=4 m_\pi^2$ is used. The 
Weinberg tree contribution to the scattering amplitude can be reduced to a constant value as
\beqa
 {\cal T}_{\rm tree}   & = &   \sum_{a=d,b=c} \left[
 \frac{1}{f_\pi^2} (s-m_\pi^2) \delta^{ac}\delta^{bd} +  
\frac{1}{f_\pi^2} (t-m_\pi^2) \delta^{ab}\delta^{cd} +
\frac{1}{f_\pi^2} (u-m_\pi^2) \delta^{ad}\delta^{bc} \right] \nonumber\\
& \Rightarrow & 
 N_\pi (2- N_\pi) \frac{m_\pi^2 }{f_\pi^2}    .
 \eeqa
The vector contribution to one loop order can be represented   as
\beqa
{\cal T}_{\rm vector}  &=&  \sum_{a=d,b=c} \left[
  \epsilon^{ace}\epsilon^{dbe} (u-t) \frac{1}{4f_\pi^4} s {\bf \Pi}_V(s)
+ 2\; {\rm permutation} \right]
\nonumber\\
&\Rightarrow &
-  N_\pi   
 \frac{s t}{2 f_\pi^4}\left[ {\bf \Pi}_V(s) +  {\bf \Pi}_V(t)\right]  
\eeqa
where 
\be
{\bf \Pi}_V (q^2) = c_1 + \frac{1}{72\pi^2} + \frac 13 \left(1-\frac{4 m_\pi^2}{q^2}\right) \left({\cal J}(q^2) - \hat c_1 \right)
\ee
and
\be
{\cal J}(q^2) = \hat c_1 +
\frac{1}{16\pi^2}\theta(q^2-4 m_\pi^2) \left( 2 + \sqrt{1-\frac{4m_\pi^2}{q^2} }\left[
\ln \left|\frac{\sqrt{1-4 m_\pi^2/q^2}-1}{\sqrt{1-4 m_\pi^2/q^2}+1}\right| + i\pi  \right] \right).
\ee
In this work, we take the mean value of the counter term $c_1=0.035$ and $\hat c_1 = 0.023$ \cite{Yam96a}.
The scalar contribution can be rewritten as
\beqa
{\cal T}_{\rm scalar}  
&=& \sum_{a=d,b=c} \left[ \frac{2 m_\pi^2}{f_\pi^4} 
\delta^{ac}\delta^{bd} \left( s {\cal J}(s)   
- \frac{5}{4} m_\pi^2  {\cal J}(s) \right)  + 2\; {\rm permutation} \right]
\nonumber\\
&\Rightarrow & \frac{2 N_\pi m_\pi^2}{f_\pi^4} 
\left(      s {\cal J}(s) + t{\cal J}(t) 
  - \frac{5}{4} m_\pi^2
\left[ {\cal J}(s) +    {\cal J} (t)  +   N_\pi   {\cal J} (0)  \right] \right) .
\eeqa
The remaining contribution can be rewritten as
\beqa
{\cal T}_{\rm rest} 
&=& \sum_{a=d,b=c} \left[ -\frac{i}{f_\pi^4} k_1^\alpha k_2^\beta p_1^\gamma p_2^\delta
\int d^4 y_1 d^4 y_2 d^4 y_3\; e^{-i k_1\cdot y_1 + i k_2\cdot y_2 - i p_1\cdot y_3} \right] \nonumber\\
&&\;\;\;\;\; \;\;\;\; \left. \vphantom{\frac 12} \times \langle 0 | \; T^*\!\!\left[{\bf j}_{A\alpha}^a (y_1) \; {\bf j}_{A\beta}^b (y_2)
\; {\bf j}_{A\gamma}^c (y_3) \; {\bf j}^d_{A\delta}(0)
\right] | 0 \rangle_{\rm conn} \right] \nonumber\\
&\Rightarrow &
\frac{N_\pi (2 + N_\pi)}{4 f_\pi^4} \left[ \vphantom{\frac 12}
(s-2 m_\pi^2)^2  {\cal J} (s) 
+ (t-2 m_\pi^2)^2  {\cal J} (t) 
+ 4 m_\pi^4 {\cal J} (0) 
\right] 
\nonumber\\
\eeqa

\end{document}